%
%% Final Version edited in Stanford, Jan 3rd, S Kachru 
\let\includefigures=\iftrue
%
% the following is to use blackboard bold fonts --
\let\useblackboard=\iftrue
%
% activate this if you don't have them.
%\let\useblackboard=\iffalse
%
% You might also need to remove this line.
\newfam\black
\input harvmac
\noblackbox
%%
%Figure Stuff
\includefigures
\message{If you do not have epsf.tex (to include figures),}
\message{change the option at the top of the tex file.}
\input epsf
\def\figin{\epsfcheck\figin}\def\figins{\epsfcheck\figins}
\def\epsfcheck{\ifx\epsfbox\UnDeFiNeD
\message{(NO epsf.tex, FIGURES WILL BE IGNORED)}
\gdef\figin##1{\vskip2in}\gdef\figins##1{\hskip.5in}% blank space instead
\else\message{(FIGURES WILL BE INCLUDED)}%
\gdef\figin##1{##1}\gdef\figins##1{##1}\fi}
\def\DefWarn#1{}
\def\figinsert{\goodbreak\midinsert}
\def\ifig#1#2#3{\DefWarn#1\xdef#1{fig.~\the\figno}
\writedef{#1\leftbracket fig.\noexpand~\the\figno}%
\figinsert\figin{\centerline{#3}}\medskip\centerline{\vbox{
\baselineskip12pt\advance\hsize by -1truein
\noindent\footnotefont{\bf Fig.~\the\figno:} #2}}
\bigskip\endinsert\global\advance\figno by1}
%%%
\else
\def\ifig#1#2#3{\xdef#1{fig.~\the\figno}
\writedef{#1\leftbracket fig.\noexpand~\the\figno}%
%\figinsert\figin{\centerline{#3}}\medskip
%\centerline{\vbox{\baselineskip12pt
%\advance\hsize by -1truein\noindent
%\footnotefont{\bf Fig.~\the\figno:} #2}}
%\bigskip\endinsert
\global\advance\figno by1}
\fi
\useblackboard
\message{If you do not have msbm (blackboard bold) fonts,}
\message{change the option at the top of the tex file.}
\font\blackboard=msbm10 scaled \magstep1
\font\blackboards=msbm7
\font\blackboardss=msbm5
\textfont\black=\blackboard
\scriptfont\black=\blackboards
\scriptscriptfont\black=\blackboardss
\def\Bbb#1{{\fam\black\relax#1}}
\else
\def\Bbb{\bf}
\fi
% *************************************
%\draft
%
\def\yboxit#1#2{\vbox{\hrule height #1 \hbox{\vrule width #1
\vbox{#2}\vrule width #1 }\hrule height #1 }}
\def\fillbox#1{\hbox to #1{\vbox to #1{\vfil}\hfil}}
\def\ybox{{\lower 1.3pt \yboxit{0.4pt}{\fillbox{8pt}}\hskip-0.2pt}}
%
%
%%MATH MACROS

\def\comments#1{}

\def\QZ{\Bbb{Z}}
\def\p{\partial}

\def\half{{1\over 2}}

\def\Im{{\rm Im\hskip0.1em}}

\def\CC{{\cal C}}

\def\CN{{\cal N}}
\def\CO{{\cal O}}%AEL

\def\CS{{\cal S}}

%AEL

\def\ap{\alpha'}

\def\II{\relax{I\kern-.10em I}}

\def\IZ{\relax\ifmmode\mathchoice
{\hbox{\cmss Z\kern-.4em Z}}{\hbox{\cmss Z\kern-.4em Z}}
{\lower.9pt\hbox{\cmsss Z\kern-.4em Z}}
{\lower1.2pt\hbox{\cmsss Z\kern-.4em Z}}
\else{\cmss Z\kern-.4emZ}\fi}
\def\IB{\relax{\rm I\kern-.18em B}}
\def\IC{{\relax\hbox{$\inbar\kern-.3em{\rm C}$}}}
\def\ID{\relax{\rm I\kern-.18em D}}
\def\IE{\relax{\rm I\kern-.18em E}}
\def\IF{\relax{\rm I\kern-.18em F}}
\def\IG{\relax\hbox{$\inbar\kern-.3em{\rm G}$}}
\def\IGa{\relax\hbox{${\rm I}\kern-.18em\Gamma$}}
\def\IH{\relax{\rm I\kern-.18em H}}
\def\II{\relax{\rm I\kern-.18em I}}
\def\IK{\relax{\rm I\kern-.18em K}}
\def\IP{\relax{\rm I\kern-.18em P}}
%\def\IX{\relax{\rm X\kern-.01em X}}
%this doesn't work

%

\def\inbar{\,\vrule height1.5ex width.4pt depth0pt}

\def\p{\partial}

\font\cmss=cmss10 
\def\IR{\relax{\rm I\kern-.18em R}}

%

 % for now

%

\def\gs{g_s}
\def\lp10{\ell_p^{10}}
\def\lp11{\ell_p^{11}}
\def\R11{R_{11}}

\def\frac#1#2{{#1 \over #2}}

%%ENGLISH MACROS

\def\cf{{\it c.f.}}
\def\etal{{\it et. al.}}
\hyphenation{Di-men-sion-al}

%%REFERENCING MACROS
\def\np{{\it Nucl. Phys.}}

\def\pl{{\it Phys. Lett.}}

\def\cmp{{\it Comm. Math. Phys.}}
\def\annm{{\it Ann. Math.}}

\lref\bdlr{I. Brunner, M.R. Douglas, A. Lawrence and
C. R\"omelsberger, ``D-branes on the Quintic'', hep-th/9906200.}
\lref\bbs{K.~Becker, M.~Becker and A.~Strominger,
``Five-branes, membranes and nonperturbative string theory,"
Nucl. Phys. {\bf B456} (1995) 130, hep-th/9507158.}
\lref\bsv{M. Bershadsky, V. Sadov and C. Vafa, ``D-branes and
Topological Field Theories,'' \np\ {\bf B463}\ (1996) 420, 
hep-th/9511222.}
\lref\thomas{R.P. Thomas, ``An obstructed bundle on a Calabi-Yau 3-fold,''
math.AG/9903034.}
\lref\ooy{H. Ooguri, Y. Oz and Z. Yin, ``D-branes
on Calabi-Yau spaces and their mirrors'',
\np\ {\bf B477} (1996) 407, hep-th/9606112.}
\lref\diacrom{D.-E. Diaconsescu and C. R\"omelsberger,
``D-branes and bundles on elliptic fibrations'',
hep-th/9910172.}
\lref\shamitjohn{S. Kachru and J. McGreevy, ``Supersymmetric
three-cycles and supersymmetry breaking'', hep-th/9908135.}
\lref\qcd{E. Witten, ``Branes and the dynamics of QCD'', Nucl. Phys. {\bf B507}
(1997) 658-690, hep-th/9706109.}

\lref\syz{A. Strominger, S.-T. Yau and E. Zaslow, ``Mirror
Symmetry is T-Duality'', \np\ {\bf B479}\ (1996) 243, hep-th/9606040.}
\lref\konthom{M. Kontsevich, ``Homological algebra of
mirror symmetry'', Proc. of the 1994 International
Congress of Mathematicians, Birkh\"auser (Boston) 1995;
alg-geom/9411018.}
\lref\daveII{D.R. Morrison, ``Mirror symmetry and
the type II string'', \np\ {\bf B}\ {\it Proc. Suppl.}\
{\bf 46} (1996) 146, hep-th/9512016.}
\lref\bundlemir{C. Vafa, ``Extending mirror conjecture
to Calabi-Yau with bundles'', hep-th/9804131.}
\lref\kmp{S. Katz, D.R. Morrison and M.R. Plesser,
``Enhanced gauge symmetry in type II string theory'',
\np\ {\bf B477} (1996) 105, hep-th/9601108.}
\lref\twopI{P. Candelas, X. de la Ossa, A. Font,
S. Katz and D.R. Morrison, ``Mirror symmetry
for two-parameter models -- I'', \np\ {\bf B416} (1994)
481, hep-th/9308083.}

\lref\kss{S. Kachru, N. Seiberg and E. Silverstein, ``SUSY Gauge Dynamics and
Singularities of 4d ${\cal N}=1$ String Vacua'', \np\ {\bf B480} (1996) 170,
hep-th/9605036\semi
S. Kachru and E. Silverstein, ``Singularities, Gauge Dynamics and Nonperturbative
Superpotentials in String Theory'', \np\ {\bf B482} (1996) 92, hep-th/9608194.}

\lref\candetal{P. Candelas, X. de la Ossa, P. Green and L. Parkes, ``A Pair of Calabi-Yau
Manifolds as an Exactly Soluble Superconformal Theory'', \np\ {\bf B359} (1991) 21.}
\lref\Fukaya{K. Fukaya, ``Floer Homology and Mirror Symmetry I,'' 1999 Kyoto preprint, and references
therein.}
\lref\wenwit{X.-G. Wen and E. Witten, ``World-sheet instantons
and the Peccei-Quinn symmetry'', \pl\ {\bf B166}\ (1986) 397.}
\lref\dsww{M. Dine, N. Seiberg, X.-G. Wen and E.
Witten, ``Non-perturbative effects on the string
worldsheet I,II'', \np\ {\bf B278}\ (1986) 769;
\np\ {\bf B289}\ (1987) 319.}
\lref\edold{E. Witten, ``New Issues in Manifolds of $SU(3)$ Holonomy'', 
\np\ {\bf B268} (1986) 79.}
\lref\distgr{J. Distler and B. Greene, ``Aspects of (2,0) String 
Compactifications'', \np\ {\bf B304} (1988) 1.}

\lref\jacques{J. Distler, ``Resurrecting $(0,2)$
compactifications'',
\pl\ {\bf 188B}\ (1987) 431.}

\lref\dkps{M.R. Douglas, D. Kabat, P.Pouliot
and S. Shenker, ``D-branes and short
distances in string theory'',
\np\ {\bf B485} (1997) 85, hep-th/9608024.}
\lref\ssmall{S. Shenker, ``Another length
scale in string theory?'', hep-th/9509132.}
\lref\witbound{E. Witten, ``Bound states of strings
and p-branes'', \np\ {\bf 460} (1996) 335,
hep-th/9510135.}

\lref\toapp{Work in progress.}

\lref\ov{H. Ooguri and C. Vafa, ``Knot
invariants and topological strings'', hep-th/9912123.}

\lref\edinst{E. Witten, ``Nonperturbative Superpotentials in String Theory'',
\np\ {\bf B474} (1996) 343, hep-th/9604030.}

\lref\seiberg{N. Seiberg, ``Electric-Magnetic Duality in Supersymmetric Non-Abelian Gauge
Theories'', \np\ {\bf B435} (1995) 129, hep-th/9411149\semi
K. Intriligator and N. Seiberg, ``Lectures on Supersymmetric Gauge Theories and Electric-Magnetic
Duality'', hep-th/9509066.}

\lref\edeva{E. Silverstein and E. Witten, ``Criteria for Conformal Invariance of (0,2) Models'',
\np\ {\bf B444} (1995) 161, hep-th/9503212.}
\lref\jdsk{J. Distler and S. Kachru, ``(0,2) Landau-Ginzburg Theory'', \np\ {\bf B413} (1994)
213, hep-th/9309110.}
\lref\edphases{E. Witten, ``Phases of N=2 Theories in Two-Dimensions'', \np\ {\bf B403} (1993) 159,
hep-th/9301042.}

\lref\mclean{R.~McLean, ``Deformations of Calibrated Submanifolds'',
Duke Univ. PhD thesis,
Duke preprint 96-01: see www.math.duke.edu/preprints/1996.html.}
\lref\hitchin{N.J.~Hitchin, ``The moduli space of special Lagrangian
submanifolds,'' dg-ga/9711002.}
\lref\joyceslag{D. Joyce, ``On counting special Lagrangian
homology 3-spheres'', hep-th/9907013.}

\lref\kodaira{K. Kodaira, ``A Theorem of Completeness of
Characteristic Systems for Analytic Families of Compact
Submanifolds of Compact Manifolds'', \annm\ {\bf 75} (1962) 146.}
\lref\katzone{S. Katz, ``On the finiteness
of rational curves on quintic threefolds'',
{\it Comp. Math.}\ {\bf 60}, 151 (1986).}
\lref\katztwo{S. Katz, ``Rational Curves
on Calabi-Yau Threefolds'', in {\it Mirror
Symmetry I} (S.-T. Yau, ed.), American
Mathematical Society and International Press (1999);
alg-geom/9312009.}
\lref\reid{M. Reid, ``Minimal Models
of Canonical 3-folds'', pp. 131-180,
Advanced Studies in Pure Mathematics 1, ed. S. Iitaka,
Kinokuniya (1983).}
\lref\burns{D. Burns, ``Some background and examples
in deformation theory,'' in {\it Complex Manifold Techniques in
Theoretical Physics}, D. Lerner and P. Sommers Eds., Pitman (1979).}
\lref\friedman{R. Friedman, ``Simultaneous resolution of threefold 
double points'', Math. Ann. 274 (1986), 671--689.}
\lref\bkl{J. Bryan, S. Katz, and N.C. Leung, ``Multiple covers and 
the integrality conjecture for rational curves in Calabi-Yau threefolds,''
math.AG/9911056.}

\lref\topsig{E. Witten, ``Topological sigma
models'', \cmp\ {\bf 118}\ (1988) 411.}
\lref\wittcs{E. Witten, ``Chern-Simons gauge theory
as a string theory'', in {\it The Floer Memorial
Volume}, H. Hofer \etal, eds., Birkhauser (1995), Boston,
hep-th/9207094.}

\lref\bpsalg{J.A. Harvey and G. Moore,
``On the algebras of BPS states'', \cmp\ {\bf 197}\
(1998) 489, hep-th/9609017.}
\lref\intth{K. Intriligator and S. Thomas, ``Dynamical
Supersymmetry Breaking on Quantum Moduli Spaces'',
\np\ {\bf B473} (1996) 121,
hep-th/9603158; ``Dual Descriptions of Supersymmetry
Breaking'', hep-th/9608046.}
\lref\fms{D. Friedan, E.J. Martinec and S.H. Shenker,
``Conformal invariance, supersymmetry and string theory'',
\np\ {\bf B271}\ (1986) 93.}
\lref\ads{J.J. Atick, L.J. Dixon and A. Sen,
``String calculation of Fayet-Iliopolous D-terms
in arbitrary supersymmetric compactifications'',
\np\ {\bf B292}\ (1987) 109.}

\Title{\vbox{\baselineskip12pt\hbox{hep-th/9912151}
\hbox{IASSNS-HEP-99/99}\hbox{SU-ITP-99/50}
\hbox{SLAC-PUB-8314}\hbox{NSF-ITP-99-142}
\hbox{OSU-M-99-13}}}
{\vbox{
\centerline{Open string instantons and superpotentials} }}
\smallskip
\centerline{Shamit Kachru$^{1,2}$, Sheldon Katz$^{3}$,}
\centerline{Albion Lawrence$^{1,2,5}$  {\it and} John McGreevy$^{4}$}
\medskip
\centerline{$^{1}${\it School of Natural Sciences,
Institute for Advanced Study, Princeton, NJ 08540}}
\centerline{$^{2}${\it Department of Physics and SLAC,
Stanford University, Stanford, CA 94305}}
\centerline{$^{3}${\it Department of Mathematics,
Oklahoma State University, Stillwater, OK 74078}}
\centerline{$^{4}${\it Department of Physics,
University of California at Berkeley,
Berkeley, CA 94720}}
\centerline{$^{5}${\it Institute for Theoretical Physics,
University of California, Santa Barbara, CA  93106}}

\bigskip
\noindent
We study the F-terms in
${\cal N}=1$ supersymmetric, $d=4$ gauge theories
arising from $D(p+3)$-branes wrapping supersymmetric 
$p$-cycles
in a Calabi-Yau threefold.
If $p$ is even
the spectrum and superpotential for a single brane 
are determined by purely classical
($\ap\to 0$) considerations.
If $p=3$, superpotentials for massless
modes are forbidden to all orders in $\ap$
and may only be generated by open string instantons.
For this latter case we find that 
such instanton effects are generically present.
Mirror symmetry relates even and odd $p$ and thus
perturbative and nonperturbative superpotentials;
we provide a preliminary discussion of a class of
examples of such mirror pairs.

\Date{December 1999}

\newsec{Introduction}

The study of D-branes in Calabi-Yau threefolds
is of both formal and phenomenological
interest. As philosophical tools, Calabi-Yau
threefolds provide a natural arena for studying
nonperturbative stringy geometry.  D-branes
are excellent probes of this geometry as they
are sensitive to structure well below the string scale
(at weak string coupling) \refs{\ssmall,\dkps}.  Furthermore
a deep understanding of mirror symmetry requires
understanding its action on D-branes \refs{\konthom,\daveII,\syz}.

The phenomenological interest is served by studying
space-filling D-branes in type I or type II
string theories, in configurations
preserving $\CN=1$ SUSY in four dimensions.\foot{The standard caveat is in
force here: in order to satisfy Gauss' law for the various RR charges,
one should either consider branes wrapping cycles in non-compact
Calabi-Yau spaces, or consider configurations containing both branes
and orientifolds.  For our purposes the former assumption will suffice, but
many of our considerations could also be applied to lower dimensional, non
space-filling branes wrapping the same cycles.}
These configurations fall into two classes \refs{\bbs,\ooy}:
6-branes in type IIA wrapping special Lagrangian cycles
of the threefold, and odd-$p$-branes in type I or
type IIB wrapping even-dimensional cycles.
The latter configurations (up to orientifolds)
can be written as coherent sheaves on the threefold
\refs{\konthom,\daveII,\bpsalg}
(or its mirror) and so involve the same type of data as
heterotic compactifications \refs{\edold,\distgr}.
The D-brane limit allows one to study gauge field
data when it is intrinsically stringy (much as Landau-Ginzburg
compactifications allow one to study intrinsically stringy
aspects of geometric data), via open string techniques.

In closed-string compactifications on Calabi-Yau
threefolds, worldsheet instanton effects are the
most well-understood source of truly stringy physics.
They drastically modify the geometry
at short distances: in addition
they lead to interesting physical effects
such as the generation of nontrivial superpotentials
in heterotic compactifications \refs{\wenwit,\dsww}.\foot{In fact,
it has been proved in \edeva\ that the most easily realized heterotic
(0,2) models, those realized as gauged linear sigma models 
\refs{\edphases,\jdsk},
are not destabilized by worldsheet instantons.  As we will 
see in the following,
it should be easier to find examples of D-brane models which exhibit disc
instanton generated superpotentials.}
In this work we will study the effects of
{\it open}-string instantons on D-brane physics,
in particular on the superpotential.  For branes
wrapping even-dimensional cycles, we will find that
the superpotential can be determined from classical
geometry; for branes wrapping special Lagrangian
cycles it is generated entirely by nonperturbative worldsheet
effects.

This has interesting implications
for mirror symmetry in the type II compactifications.
To begin with, if a given mirror pair
of cycles has massless deformations with a nontrivial
superpotential, then the classical moduli spaces 
will not match under the mirror map.  A holomorphic
2-cycle with its infinitesimal holomorphic deformations
obstructed at some nontrivial order
(so that a 5-brane wrapped around it will have
massless chiral fields with a superpotential)
will have as its mirror a 3-cycle which has flat directions
to all orders in $\ap$.  On the other hand, if we
start with a special Lagrangian cycle and find that
worldsheet instantons destabilize or make nonsupersymmetric
the D-branes wrapping them,  the mirror will respectively
either not exist or will be
some classically nonsupersymmetric configuration.  This is
reminiscent of a common feature of dualities of
${\cal N}=1$
gauge theories, where superpotentials generated by nonperturbative dynamics
are dual to tree-level superpotentials \seiberg.
However, it is a relatively novel situation for ${\cal N}=1$ dualities 
of string vacua, where normally instanton effects map to instanton effects, as
in heterotic/F-theory duality \edinst.  Here mirror
symmetry should provide a powerful tool for
summing open string instantons, as it does for closed
string worldsheet instantons \candetal. 
Note that the nonperturbative superpotentials we are
discussing here are not explicable in terms of gauge dynamics involving
the (perturbative) D-brane gauge theory; we would expect in some circumstances
the instanton effects will be related to gauge instantons 
of a $\it {non-perturbative}~$
D-brane gauge theory that arises at singular points 
in the brane moduli space.
This is analogous to the fact that in heterotic (0,2) 
models, worldsheet instantons
can sometimes be related to gauge instantons of 
nonperturbative gauge groups arising
from singular compactifications \kss.

In this paper we will begin to investigate disc instanton effects
by asking whether nonperturbative superpotentials
are generically generated by worldsheet (disc) instantons. We will
find a story similar to that of the heterotic string
\dsww: when the open string instantons
are isolated, nonvanishing (locally) runaway
potentials may be generated.\foot{However, in contrast to the
heterotic string story, the classical moduli spaces of the brane
configurations we will study are naturally compact.  Hence, the
superpotentials we find will have minima which are not
``at infinity'' in field space.}

The plan of our paper is as follows: in \S2
we will review the construction of $\CN=1$ four-dimensional
theories via space-filling D-branes wrapped on Calabi-Yau
threefolds in type II string theory.
After discussing
the results determined by classical geometry,
we will discuss some constraints from string theory.
First, at tree level the superpotentials are
computable via topological open string theory, and
the hypermultiplets of the background closed-string theory
decouple \bdlr.  Furthermore, we will find that in the case
of D6-branes wrapping special Lagrangian three-cycles,
superpotentials are forbidden not only classically
but to all orders in $\ap$, due to
a Peccei-Quinn symmetry.  Superpotentials in these
cases can only be generated by topologically nontrivial
disc instanton effects.  In \S3 we will
discuss the generic superpotential terms that
are allowed in the presence of an isolated holomorphic
disc.
Finally, in \S4 we
discuss promising directions for future work.  Further results in explicit
examples
will appear in a companion paper \toapp.

There is a close relation between the ideas discussed in this paper and
earlier work of Witten \wittcs\ and Vafa \bundlemir.  As this work was being
completed, we were also informed of the related work 
\Fukaya\ by mathematicians.

\newsec{Classical geometry of D-branes on $CY_3$}

We begin with D-brane configurations preserving
four supercharges, to lowest order in
$\gs$ and $\ap$, in type II string theory
on $M\times\IR^4$ where $M$ is
a Calabi-Yau threefold.
We will assume that the D-branes fill all of spacetime so that we
realize an $\CN=1$, $d=4$ gauge theory.  

The internal configurations
preserving four supersymmetries fall
into two classes \refs{\bbs,\ooy}: ``A-type'' branes
wrapping special Lagrangian submanifolds of the threefold,
and ``B-type'' branes which wrap holomorphic cycles
of the Calabi-Yau.  (The latter may have nontrivial holomorphic
gauge bundles living on them as well, corresponding
to bound states with lower-dimensional branes.
We will for the most part ignore this possibility.)
In the present discussion these exist in the type IIA and IIB
theories respectively.
We will discuss the associated gauge theories
of each class in turn.

\subsec{A-type branes}

\bigskip\noindent{\it Spectrum}

In the large-volume, large-complex-structure
limit, supersymmetric A-type branes 
wrap special Lagrangian submanifolds.
Let such a manifold $\Sigma$ be described by a map
$$ f: \Sigma \to M\ . $$
Recall that special Lagrangian submanifolds $\Sigma$
are defined by the properties that:
\eqn\slag{
\eqalign{
        \dim_\IR \Sigma &= \half \dim_\IR M \cr
        f^\ast \omega &= 0 \cr
        f^\ast (\Im e^{i\theta}\Omega) &= 0
}}
where $\omega$ is the K\"ahler form of $M$,
$\Omega$ is the standard holomorphic $(3,0)$ form,
and $e^{i\theta}$ is some phase.

$N$ D6-branes wrapping a single supersymmetric
cycle $\Sigma \subset M$
have a $U(N)$ vector multiplet arising from
massless open string excitations
polarized completely in $\IR^4$.
Massless open string excitations polarized in $M$
form adjoint $U(N)$ chiral multiplets.  We will focus
on the case $N=1$.  To lowest order in $\ap$, the counting
of massless chiral fields has been worked out.
$\Sigma$ lives in a family of deformations
with dimension $b_1(\Sigma)$ \mclean\ (\cf\ also
\hitchin\ for a clear discussion).  More
precisely, each basis vector in the tangent space
to the space of deformations
may be used to construct a nontrivial
harmonic one-form on $\Sigma$, and vice-versa.
Of course the space of such deformations (which has
real dimension $b_{1}(\Sigma)$) cannot
make up our set of chiral multiplets which are built
from complex scalars: for example
$b_1(\Sigma)$ need not be even.  However,
deformations of flat connections of the D6-brane gauge field
on $\Sigma$ also map one-to-one onto the space of
harmonic one-forms on $\Sigma$, roughly because there is
a Wilson line of the $U(1)$ gauge field around each 1-cycle. Thus
for each element of the space $\CH^1(\Sigma)$ of
harmonic one-forms on $\Sigma$ one has two real
flat directions which may be described by a complex
scalar \syz.  In other words, we find 
$b_1(\Sigma)$ massless chiral multiplets,
one for each non-trivial one-cycle or
harmonic one-form on $\Sigma$.

Note that if we have branes wrapping several
(mutually supersymmetric) 3-cycles, then
we may get additional matter from any intersection
points, in bifundamentals of the $U(1)$s of each
cycle.  A local example of this was
discussed in \shamitjohn.  In this work we will discuss branes
wrapping single ``primitive'' 3-cycles: however,
as explained in \shamitjohn, interesting transitions
to this more complicated case can occur as one varies
background (closed string) hypermultiplets.

A natural choice of coordinates on the moduli space of
the wrapped D6 brane is the following \refs{\bundlemir,\joyceslag}.
Let $\{\gamma_j\}$ be a basis for ${\CH}_{1}(\Sigma)$.
Choose minimal area discs $D_j$ subject to the condition that
\eqn\boundary{\partial D_j ~=~\gamma_j}
and let
\eqn\area{w_j = \int_{D_j} \omega~.}
In other words, if there is a holomorphic disc
in the relative homology class of $D$, then $w_j$ will be the area.
The $w_j$ provide $b_{1}(\Sigma)$ real coordinates.
They are complexified by $b_1(\Sigma)$ Wilson lines
\eqn\wilson{a_j ~=~ \int_{\gamma_j} A}
where $A$ is the $U(1)$ gauge field on the wrapped brane.
The coordinates in \area\ and \wilson\ are the real and imaginary
parts of scalar components of the $b_1(\Sigma)$ chiral multiplets
$\Phi_j$ on the brane:
\eqn\supermul{\Phi_{j} = w_j + i a_j + \cdots}
where $\cdots$ indicates higher components of the superfields.

\bigskip\noindent{\it Superpotentials and worldsheet instantons}

As noted above, it follows from McLean's theorem \mclean\ that
at lowest order in $\alpha^\prime$ the brane wrapping $\Sigma$
has a moduli space of dimension $b_{1}(\Sigma)$.
So far there may still be $\ap$ corrections
which lift these flat directions.
To lift moduli, one would need to generate either D-terms or
F-terms in the low energy action; we will focus on the superpotential,
since the chiral multiplets $\Phi_j$ are neutral (at least at generic points
in the classical moduli space of the cycle) and do not appear in
FI D-terms.  At leading order in $\ap$, the superpotential $W(\Phi)$
identically vanishes.  We now determine to what
extent $\ap$ corrections of any sort are possible.

In fact, it turns out that there are
no corrections to the open-string superpotential
to any finite order in $\ap$:
all contributions must come from
nonperturbative corrections, arising
from topologically nontrivial configurations.
The arguments are almost identical to similar arguments
for the heterotic string \refs{\wenwit,\dsww}.  We
will give two.

The first argument is a string theory argument.
The $(0)$-picture vertex operator 
for a flat connection $A$ 
at zero momentum 
on the
D6-brane has the form:
\eqn\wilsonvert{
        V = \int_{\p D} A_\mu(X)\p_\alpha X^\mu d\sigma^\alpha
}
where $X$ are coordinates on the brane and $\sigma$
coordinates on the worldsheet $D$.  Let $A$ be polarized completely
internally, so that it corresponds to a choice of Wilson
lines around the elements of $H_1(\Sigma)$.  If $X(\p D)$
is a topologically trivial cycle on $\Sigma$, then $A$
can be written as an exact form $d\Lambda$.  We see that $V$ vanishes
after an integration by parts.  Thus
topologically trivial disc amplitudes give no non-derivative
couplings (such as superpotential terms) of
the imaginary parts of chiral multiplets, to all orders
in $\ap$.  Holomorphy thus requires the superpotential
vanish to all finite orders in $\ap$.

If the boundary maps
to a topologically {\it non}-trivial cycle
$\gamma_j \subset \Sigma$, this argument fails.
Such discs are non-trivial elements of the
relative homology class $H_2(M,\Sigma)$.
These worldsheets will give terms weighted
by the instanton action:
\eqn\instweight{
        e^{-(w_j + i a_j)/\ap}
}
where $w_j$ is the spacetime area of this disc as in \area, 
and $a_j$ is its partner Wilson line
\wilson.
The $\ap$ dependence is decidedly non-perturbative.
Note that to obtain the contribution \instweight\ to the
action, the map $X(\sigma)$ must be a $\it holomorphic$ map from
the disc to $M$ with the desired boundary, and with the normal derivative to
$X(\sigma)$ at the boundary taking values in the 
pullback of the normal bundle to
$\Sigma$; this is a
disc instanton.  It follows from standard arguments that
only such holomorphic maps have the correct
zero mode count to contribute to a superpotential term in
the spacetime theory.

The second argument is a spacetime argument.
The space of
$U(1)$ Wilson lines on a circle is
the dual circle.  Thus,
any function of the chiral fields 
appearing in the effective action 
must be invariant
under discrete shifts of their imaginary
parts.  Holomorphy then requires that the
superpotential be a power series in
$\exp\left(-[w_j + i a_j]/\ap\right)$,
which again gives a nonperturbative dependence
on $\ap$.

\bigskip\noindent{\it Examples of three-cycles}

The best-known
example arises in the Strominger-Yau-Zaslow formulation
of mirror symmetry \syz: the claim is that any geometric
Calabi-Yau with a geometric mirror can
be written as a fibration of special Lagrangian
$T^3$s.  Mirror symmetry is fiberwise T-duality on these
$T^3$s.  D3-branes wrapping these fibers are mapped
to D0-branes on the mirror.  The $T^3$ has $b_1 = 3$
so all this is in accord with expectations: the
mirror D0-brane and thus the wrapped D3-brane should have
a 3-dimensional complex moduli space (which is the mirror
threefold).
Many examples of 
special Lagrangian three-cycles can be
found as fixed loci of real structures.  Some examples, 
which are homeomorphic to $\IR\IP^3$, 
are contained in \refs{\bbs,\bdlr}.
Note that these have a
$\QZ_2$ Wilson line degree of freedom as
$\pi_1(\IR\IP^3)=\QZ_2$.

In addition, there has been some discussion of
local and non-compact models.  Ref. \joyceslag\
contains some general discussion of noncompact
supersymmetric three-cycles.
A simple example with an isolated disc instanton is
the following.  Take $z_{1,2,3}$ as coordinates on
$\IC^3$, and choose $\omega$ and $\Omega$ to be the
obvious K\"ahler form and holomorphic three-form.
Then the three-cycle $\Sigma$ defined by
\eqn\eg{|z_1|^2 - t = |z_2|^2 = |z_3|^2}
\eqn\egtwo{Im (z_1z_2z_3) = 0, ~~Re (z_1z_2z_3) \geq 0}
with $t$ positive is special Lagrangian, and diffeomorphic to
$S^1 \times \IR^2$.
A generator $\gamma$ of $H_{1}(\Sigma)$ is given by the concrete
choice
\eqn\gener{\gamma: \{(t^{1/2} e^{i\theta},0,0)\}}
where $\theta$ runs from 0 to $2\pi$.
A holomorphic disc with boundary $\gamma$ and area
$\pi t$ is given by
\eqn\disc{D_{t}: \{(z_1,0,0), ~|z_1|^2 \leq t)\}}

\subsec{B-type branes}

In the large-volume, large-complex-structure limit,
supersymmetric B-type branes wrap holomorphic cycles of $M$.
One may also examine bound states which can
be described as gauge bundles on the highest-dimensional
branes (\cf\ for example \bdlr\ for a discussion). 
For simplicity we will focus on branes
wrapping primitive cycles, and in our examples we will
discuss cases where $\CC$ is a rational curve.

For $N$ (space-filling) D-branes wrapping a given cycle $\CC \subset M$
one again has
a $U(N)$ vector multiplet arising from massless open strings
polarized along the spacetime directions.  Massless
strings polarized along $M$ give rise
to adjoint chiral multiplets.  Again we will focus
on $N=1$. For B-type branes
wrapping $\CC$ the infinitesimal
supersymmetric deformations of the cycle
are holomorphic sections of the normal
bundle $\CN_\CC$.  The number of
such first-order deformations is therefore 
the dimension of the
space of holomorphic
sections, $H^0(\CC, \CN_\CC)$
(this is the cohomology group of the bundle $\CN_\CC$,
not a relative cohomology group).  These
are the scalars in the massless
chiral multiplets.  There is no guarantee
that these deformations do not have an obstruction at 
higher order.\foot{For e.g. $\CC$ a curve of genus $g \geq 1$, there
are also $2g$ Wilson line degrees of freedom which parametrize the
flat $U(1)$ bundles on $\CC$.  These pair up into
$g$ chiral multiplets and provide exactly flat directions.
Similar comments apply if $\CC$ is a four-cycle with 
$b_{1}(\CC) \neq 0$.}  
Such obstructions, if they
exist, correspond to elements of
the group $H^1(\CC, \CN_\CC)$ \kodaira.
More specifically, given an element of
the cohomology group
$H^0(\CC, \CN_\CC)$ one may try
to construct a finite deformation by beginning
with an infinitesimal deformation and constructing
a finite deformation as a power series.
$H^1$ measures the space
of possible obstructions at each order in this series.  Note
that it may happen that although $H^1$ is nonvanishing,
there is still a solution for this power series and
thus a family of cycles.  In the end, an obstruction
should appear as a higher-order term in the superpotential
for a brane wrapped around this cycle \bdlr.

Furthermore, deforming the complex structure of $M$
can also cause obstructions to (previously existing)
deformations of $\CC$.
The basic statement is as follows
(\cf\ \refs{\burns,\twopI}).
One may use the restriction map to $\CC$ and
the short exact sequence:
\eqn\shortembed{
	0 \to T_\CC \to T_M |_\CC \to \CN_{\CC} \to 0
}
to write a map
\eqn\rmap{r: H^1(M, T_M) \to H^1(\CC,\CN_{\CC}) \ .}
If we perturb the complex structure of $M$ to first order by some element
$\rho \in H^1(M,T_M)$, a deformation of $\CC$ exists
which preserves $\CC$ as a holomorphic cycle
if and only if $r(\rho) = 0$.  Note that couplings
of (open-string) chiral multiplets
to (background closed-string)
complex structure parameters in the superpotential
are allowed and generic \bdlr.

In the end, even counting
these chiral multiplets is a harder problem on its face
than for A-type branes, as the number of moduli depends
not only on the intrinsic topology of the cycle but
on the details of its embedding in $M$.  (This
is already apparent for rational curves in the quintic --
\cf\ \refs{\katzone,\katztwo}.)  Nonetheless, one may find a lot
of specific examples for which computations are
possible, especially for rational curves.

Some additional constraints exist as
for A-type branes.  First, the computation of the
disc contribution to the superpotential can be reduced
to a B-twisted open topological field theory calculation
\bdlr.  Again the K\"ahler parameters almost completely
decouple from the superpotential: indeed, the
computations of the dimension of $H^0(\CC, \CN_\CC)$
and of the obstruction depend completely on the complex
structure.  But in addition,
all contributions to B-model computations come entirely
from constant maps into the target space \refs{\topsig,\wittcs}.
There are no worldsheet instanton corrections
and tree level sigma model calculations will suffice:
the superpotential can be deduced from classical
geometry.

\vfill\eject
\bigskip\noindent{\it Examples of holomorphic curves and superpotentials}

Many useful examples of holomorphic cycles exist
in the literature.  Several can be found or
are referenced in \bdlr.  We are particularly interested
in two-cycles with nontrivial obstructed deformations.

The canonical example is simply a small resolution
of the singular hypersurface in $\IC^4$:
\eqn\localmodel{
        xy = z^2 - t^{2n}
}
(Such a small resolution is consistent with the
Calabi-Yau condition as the space is noncompact.)  If $n=1$,
then $H^0(\CC,\CN_\CC)=0$, and the curve is rigid.  If $n>1$, then
$H^0(\CC,\CN_\CC)$ is one-dimensional ---
the normal bundle to this curve is $\CO(0) \oplus\CO(-2)$ ---
but there is an obstruction at $n$th order to
deforming this curve \reid.  This phenomenon
can be described by a superpotential
$W(\Phi)=\Phi^{n+1}$ \bdlr.

It is easy to use $W(\Phi)$ to see the effect of a general deformation
of complex structure on $C$.  We perturb $W(\Phi)$ to
\eqn\npert
{W_t(\Phi)=\Phi^{n+1}+tP(\Phi)+O(t^2),}
where $P(\Phi)$ is an arbitrary
polynomial in $\Phi$ subject only to the genericity condition
$P'(0)\ne 0$.  Solving $W_t'(\phi)=0$, we get $n$ solutions for the vev
$$\phi_k(t) = e^{2\pi i k/n}\left(-{P'(0)\over n+1}\right)^{1/n}
     t^{1/n}+\ldots,$$
where the dots denote higher order terms in $t$.
The geometric description of this perturbation of curves with
normal bundle $\CO(0) \oplus\CO(-2)$ to $n$ rigid
rational curves was well known \friedman.  The geometric perturbation of
contractible curves with normal bundle $\CO(1)\oplus\CO(-3)$ to rigid
curves has recently been worked out in
\bkl\ and can be rephrased in terms of the perturbation of a superpotential if
desired.  The geometric description in the case of a general $\CO(1)\oplus
\CO(-3)$ curve is not yet worked out, but the introduction of a superpotential can
be expected to clarify the geometry.

\bigskip\noindent{\it Digression on holomorphic Chern-Simons theory}

Another way to arrive at the superpotential in 
examples like \localmodel, \npert\  
is by studying a holomorphic analogue of the Chern-Simons action, discussed
in \refs{\wittcs,\bundlemir}.\foot{We thank
C.~Vafa for pointing this out to us, and D. Diaconescu for 
related discussions.} 
In the following, we will suppress constants which enter
harmlessly in our formulas.
We think of the Calabi-Yau
$M$ as being obtained from the total space 
of the normal bundle $\CO(0) \oplus\CO(-2)$ by a modification
of the complex structure.  We choose holomorphic coordinates $(z,z_0,z_1)$ on
the normal bundle, with $z$ being a coordinate on $\CC$, $z_0$ being in 
$\CO(0)$, and $z_1$ in $\CO(-2)$.  The curve $\CC$ is identified with
the zero section $z_0=z_1=0$.  The modification of complex structure is
realized as usual by perturbing
$\bar\partial$ by a tensor $A^i_{\bar{j}}$, i.e.\ 
$\bar\partial_{\bar{j}}\mapsto
\bar\partial_{\bar{j}}+A^i_{\bar{j}}\partial_i$, where 
$A^{i}_{\bar j}$ is a $TM$ valued $(0,1)$ form on $M$.  We assume that 
the curve $\CC$
remains holomorphic, and want to understand which deformations $z_i=\phi_i(z)
\ (i=0,1)$ remain holomorphic.  The space of $C^\infty$ deformations of
$\CC$ is identified with the space $(\phi_0,\phi_1)$ of $C^\infty$ sections
of the normal bundle $\CN_{\CC}$.  The relevant holomorphic 
Chern-Simons action is
\eqn\cs
{\int_{\CC} \left(\phi_0\left(\bar{\partial}+A^i_{\bar{z}}\partial_i\right)\phi_1 -
  \phi_1\left(\bar{\partial}+A^i_{\bar{z}}\partial_i\right)\phi_0\right).}
Note that in \cs\ we only use the index $\bar{j}=\bar{z}$ in $A$.
\cs\ expands as
\eqn\csexp
{\int_{\CC} 
\phi_0\bar\partial\phi_1+\phi_0A^z_{\bar{z}}\partial\phi_1+\phi_0A^1_{\bar{z}}
-\left(
\phi_1\bar\partial\phi_0+\phi_1A^z_{\bar{z}}\partial\phi_0+\phi_1A^0_{\bar{z}}
\right).}
We make sense of this by respectively identifying $\phi_0,\phi_1$ 
with functions and $(1,0)$ forms on $\CC$ (as would be expected
in the twisted brane worldvolume theory \bsv), while respectively
identifying $A_{\bar{z}}^0,A_{\bar{z}}^1$ with $(0,1)$ and $(1,1)$ forms
after pulling back to $\CC$.
Thus all the terms in \csexp\ are $(1,1)$ forms on $\CC$ and can be integrated.

The variations of \cs\ or \csexp\ with respect to $\phi_0$ and $\phi_1$ give
the conditions that the corresponding curve in $X$ is holomorphic.  In fact, 
the action of the topological theory on 
$\CC$ actually becomes the superpotential in the four-dimensional
${\cal N}=1$ theory arising from  
wrapping a D5 brane on $\CC$.
This is because the holomorphic Chern-Simons theory is the
string field theory for the open string topological B-model \wittcs, and 
therefore
its action is the generating function of the topological correlation
functions which give rise to superpotential terms in the physical theory.

To illustrate this fact, we now show that we can choose our tensor $A$ so that
\csexp\ becomes $W(\Phi)=
\Phi^{n+1}$.  Since the obstructions to deforming $\CC$ lie in $H^1(\CN_{\CC})
=H^1(\CO(-2))$, we choose our $A$ to have $A_{\bar{z}}^1 = z_0^n dz\wedge
d\bar{z}$ while the other $A^i_{\bar{z}}$ vanish (we can always choose
such a gauge).  Then the constant section
$(\phi_0,\phi_1)=(t,0)$ is holomorphic provided we put $t^n=0$.  So this
$A$ produces the required geometry. 

The variation of \csexp\ with respect to $\phi_1$ shows that $\phi_0$ is 
holomorphic.  The variation of \csexp\ with respect to $\phi_0$ shows
that $\bar{\partial}\phi_1$ is a multiple of $\phi_0^n$.  Substituting
these back into \csexp\ (and performing the integral over the curve
$\CC$, which just produces a volume factor) 
gives a multiple of $\phi_{0}^{n+1}$, as claimed.
This proves that for any $\CO(0) \oplus \CO(-2)$ curve, the superpotential
will be a polynomial of degree $k$ for some $k$ (or will vanish identically)
-- $k$ is the only invariant of the complex structure in some neighborhood
of the curve.

\bigskip\noindent{\it Another Example} 

Another example which we will use was detailed in
Ref. \kmp\ (see also sec. 9 of \twopI.)  Here one
has at a specific point in the complex structure
moduli space an $A_1$ singularity fibered over a
genus-g curve $\CS$.  At this point the collapsing
cycles obviously form a family which is precisely $\CS$.
Deformations of the
complex structure of $M$ destroy this family, generically 
leaving $2g-2$ isolated curves.
One may find $2g$ three-cycles by sweeping the
collapsing curves over the one-cycles
of $\CS$, mapping $H_1(\CS)$ to $H_3(M)$.\foot{This is closely related
to the formula for the superpotential in 
\qcd.  Fixing a point $s_0\in\CS$, then a path from $s_0$ to $s\in\CS$ sweeps
out a 3-chain in $M$, which can be integrated over a holomorphic
3-form, defining a function of $s$.  If we define the potential this
way in our context, there is a multiplicative ambiguity from the
choice of holomorphic 3-form, reflected in the description in the main
text by the choice of isomorphism $H^{(2,1)}(M) \simeq H^1(M, T_M)$.}
This can be lifted to a map from $H^{(1,0)}(\CS)$ into $H^{(2,1)}(M)
\simeq H^1(M, T_M)$
\refs{\kmp,\twopI}.  This gives $g$ independent first-order deformations of
complex structure.\foot{If $M$ arises from Batyrev's construction
of Calabi-Yau toric hypersurfaces by blowing up the curve $\CS$ of $A_1$
singularities, these deformations of the complex
structure of $M$ are those which are not realizable
by polynomial/toric deformations.}  We can use the map $r$
\rmap\ to project the relevant deformation onto $H^1(\CN)$
for each fiber of this collapsing surface.
Now the spaces $H^1(\CN)$ are the fibers of a bundle over $\CS$, and this
bundle is identified with the canonical bundle of $\CS$.
So \rmap\ gets included in the sequence of maps
\eqn\obst{
H^{(1,0)}(\CS) \to H^1(M,T_M) \to H^0(\CS, K_\CS).}
This says two things.  First of all, first order differentials on $\CS$
lead to first order deformations of complex structure, 
realizing $g$ deformations of complex structure.
Second, upon perturbing by such a complex structure deformation,
the only curves which survive the deformation are those which are located at
zeros of the associated section of $K_\CS$.  Thus generally
we will find a set of isolated curves with only massive chiral
multiplets.  However, at codimension one in the complex
structure moduli space, zeros of the section of
$K_\CS$ will coincide and the resulting curve will have higher
multiplicity: their deformations will be massless
but obstructed at some higher, non-trivial order.

A superpotential which reflects this geometry be constructed as follows.
Let the complex structure deformation be induced as above from
an element $\omega \in H^{(1,0)}(S)$.  At $\omega=0$ the
curve lives in a family which is
precisely $\CS$ so that a deformation of a curve at $z\in \CS$
described by $\CN$
can also be written as an element $\phi\in T_z^{(1,0)}\CS$.
One may then write the superpotential as:
\eqn\superp{W(\Phi;\omega) ~=~
	\langle \omega,\Phi \rangle +
	{1\over 2!}\langle \p \langle \omega,\Phi \rangle ,\Phi\rangle +
	{1\over 3!}\langle \p \langle \p \langle \omega,\Phi \rangle ,
                \Phi\rangle, \Phi\rangle + \cdots
}
Here $\Phi$ is the superfield associated to $\phi$, $\p$ is 
the the Dolbeault operator
on $\CS$ and
$\langle,\rangle$ is the usual inner product between forms
and vectors.  It is understood that one is to evaluate the 
inner product at the
point $p \in \CS$ around which one is expanding, and convergence follows
from the convergence of the power series representation of $\omega$.
For $\CS$ of genus $g$, the
expansion \superp\ can be truncated after $2g-1$ terms without changing
the location and structure of the critical points.
The closed string complex structure moduli act as parameters in
the superpotential, through the choice of $\omega$.

Let us explore the properties of \superp\ slightly more explicitly, 
to illustrate
its features.  Consider expanding \superp\ about some point on $\CS$ where
$\omega$ has an expansion in a local complex coordinate $z$
\eqn\omegaexp{\omega ~\sim~ z^{n} dz.}
We can represent the scalar field, which we are thinking
of as a tangent vector to $\CS$, as 
$\phi {\partial\over {\partial z}}$ with $\phi$ complex.
Then, expanding \superp\ around $z=0$, we find

$$W(\Phi) ~\sim~ \Phi^{n+1}.$$

\noindent
For $n=0$ (i.e. around generic points on $\CS$) there is no
supersymmetric vacuum, while for $n>0$ there are supersymmetric vacua.
For $n=1$, the vacuum is massive; for $n>1$ there is a massless field,
and the vacuum splits into $n-1$ massive vacua upon small
perturbations of the complex structure of $M$ (just as in the
situation of \npert).  For $\CS$ of genus $g$, $\omega$ will
generically have $2g-2$ isolated zeroes, giving rise to $2g-2$ massive
supersymmetric vacua at generic points in the space of background
closed string parameters.  At various codimensions in the closed
string moduli space, as one further specializes the multiplicities of
the zeroes of $\omega$, these $2g-2$ vacua merge in various
combinations to yield theories with massless fields obstructed by
higher order potentials.

A simpler way to write \superp\ locally on $\CS$ is to write $\omega=
df_\omega$ for a locally defined function on $\CS$.  Locally, such an
$f$ can be thought of as a function of $\phi$.  Then we simply have
$$W(\Phi,\omega)=f_\omega(\Phi).$$
While this formula is simpler in form than \superp, it does not capture
the global structure of the moduli space $\CS$.

The above considerations are easily adapted to the more general
situation considered in \kmp, where an $A_N$ singularity is fibered
over $\CS$.  If we denote the collapsing curve as $C_1\cup\ldots\cup
C_N$, then for each $C_j$ we get $g$ deformations of complex structure
arising as in \obst, yielding $gN$ complex moduli.  But we also have
connected subsets $C_k\cup C_{k+1}\cup\ldots\cup C_{k+r}$ to which the
above analysis applies.  But since the first map in \obst\ depends
linearly on the individual $C_j$, including these connected subsets
does not give rise to any new complex structure deformations.  So we
get $N(N+1)/2$ superpotentials of the form \superp\ on $N(N+1)/2$
copies of $\CS$, each of which depends on the $gN$ complex moduli (only $g$
of which appear in any one superpotential).  
Each of these superpotentials controls the obstructions to deforming
curves of the form $C_k\cup C_{k+1}\cup\ldots\cup C_{k+r}$, and $2g-2$
curves of this type survive a generic deformation of complex structure.

\newsec{Disc instantons}

Type II string theory in the presence of a D-brane on a given special
Lagrangian submanifold
has the same net number of
worldsheet (and spacetime) supersymmetries as a heterotic
$(0,2)$ model; and as with heterotic (0,2) models,
the nonrenormalization
theorem for the spacetime superpotential is
spoiled by worldsheet instanton effects.
In light of results for $(0,2)$ models \dsww, it is fair to
ask whether the generic D6-brane configuration is 
nonperturbatively stable.

We expect direct calculations of instanton effects
to be difficult.
But instantons in supersymmetric theories
generate fermion zero modes which provide selection
rules for CFT correlators.  Using the
rule of thumb that allowed terms are generic, we
will see that three-cycles with an isolated disc instanton
are destabilized nonperturbatively.\foot{Holomorphic
discs ending on special Lagrangian cycles of
Calabi-Yau threefolds are generically isolated \wittcs.}
The argument is quite similar to that for heterotic
$(0,2)$ models.

The easiest way to count the zero modes for an
isolated holomorphic
disc is to begin with the amplitude for the
sphere and get the disc by orbifolding with respect to
a real involution, which will cut the number of
zero modes in half.
For an isolated sphere, the superconformal symmetry
together with an index theorem shows that there are four
holomorphic zero modes and four antiholomorphic zero modes
\refs{\dsww,\jacques}, so
we expect four fermion zero modes on the disc.

Consider a single D6-brane wrapping 
a special Lagrangian three-cycle $\Sigma$.
The complex modulus $\phi = w + ia$ 
is associated with a 
cycle $\gamma \in H_1(\Sigma)$, using the
notation and definitions of 
\S2.1.
Here we assume the isolated instanton
corresponds to a disc $D$ such that
$\p D = \gamma$ and $D$ has minimal area.
The most obvious, lowest-order term consistent
with our perturbative nonrenormalization theorem
is the exponential
\eqn\simplesuper{
	W(\Phi) = e^{-\Phi/\ap}
}
where $\Phi$ is the superfield
corresponding to $\phi$.  This will
clearly destabilize the wrapped D6-brane, at least
locally.

We will search for the superpotential $\simplesuper$
by examining small
fluctuations $\Phi_j$ away from the above
classical configuration
$\Phi_0 = \phi$.  Here $j$ is an index
in $H_1(\Sigma)$.
The lowest-order terms directly computable via
a CFT correlator will be those arising from
the cubic term
$$ \Phi_i\Phi_j\Phi_k e^{-\phi/\ap} \ . $$
We will focus on the term
\eqn\phicubed{
	S_{{\rm cubic}} = C \int d^4x \phi_i\phi_j F_k
}
where $F_k$ is the auxiliary field in $\Phi_k$.  
Note that in the
reduction to four dimensions, the operators
above are arrived at by contour integrals
in $\Sigma$, so $C$ is proportional to a triple
integral.

The vertex operators which enter in the 
calculation of \phicubed\ are easily presented
in the covariant RNS formalism \fms\ 
(\cf\ \bdlr\ for a general discussion
of the CFT calculation of open-string
superpotential terms).  The $(-1)$-picture
zero-momentum vertex operator for the
scalar component $\phi_j$ is:
\eqn\scalvert{
	V^{(-1),j}_{\phi} = \theta^{j}_\mu(X)\psi^\mu 
	e^{-\tilde{\phi}}
}
where $\tilde{\phi}$ is the bosonized
superconformal ghost \fms,
$\theta^{j}_\mu$ is the harmonic one-form
(associated to $\gamma_j$) on the 3-cycle, and $\psi^\mu$ is a
fermion with Dirichlet boundary conditions.
The $(0)$-picture vertex operator
for the auxiliary field is \ads:
\eqn\auxvert{
	V^{(0),j}_F = \Omega_{\rho\mu\nu}(X)\theta^j_\sigma 
		g^{\sigma\rho}\psi^\mu \psi^\nu\ .
}
Here $\Omega$ is the the $(3,0)$ form, with the coordinates
(but not the indices) restricted to $\Sigma$.
Equation \auxvert\ is obtained by applying the unit spectral flow operator
$\Omega_{\mu\nu\rho}\psi^\mu\psi^\nu\psi^\rho$ as in 
\refs{\ads,\distgr}.

The three-point function
\eqn\threept{\langle V^{(0),i}_\phi 
	V^{(-1),j}_\phi V^{(-1),k}_F \rangle}
has the correct fermion and ghost number to
be nonvanishing; in an instanton background,
the four fermions in 
the vertex operators in \threept\ can soak up the
relevant zero modes.  Note that we are computing
the integrand of the triple integral defining $C$
in Eq. \phicubed.  Since holomorphic maps
will preserve the order of marked points on
the boundary, the ordering of \threept\ will
be fixed for a given set of positions in this
integrand.

This superpotential term can equivalently be computed as
a correlator in the topological A-model open string
theory \refs{\wittcs,\bdlr}.  Here one
is computing the contribution to \threept\ (or more familiarly, a Yukawa
coupling related to \threept\ by supersymmetry) in a
sector where the map of the worldsheet to spacetime is
a disc whose boundary $\gamma \subset \Sigma$
is topologically nontrivial.  The path integral
localizes onto the space of holomorphic maps, and
the contribution is 
\eqn\threeint{
	\oint_\gamma\oint_\gamma\oint_\gamma 
		dx_1 dx_2 dx_3 A_i (x_1) A_j(x_2) A_k(x_3)
}
(suppressed by the exponential of the 
area of the holomorphic disc), where the gauge fields
$A_i$ can be identified with the 1-forms $\theta^i$ in \scalvert. 
Once again, for given positions in the
integrand, the ordering of the vertex
operators for $A_{i,j,k}$ must match the
ordering of $x_{1,2,3}$ respectively.

The result is that the superpotential \simplesuper\
is generic for an isolated instanton.
With some interpretation added, this
statement matches a calculation in ref. \wittcs.
There it is shown that the string field theory
for the topological open string A-model
is equivalent to Chern-Simons theory on $\Sigma$
with instanton corrections to the action.
This instanton correction can be interpreted as
precisely the superpotential we have calculated,
as it generates the topological correlator we have discussed.
Note that in \wittcs, the dependence
on the area of the disc was added as a convergence factor, 
whereas in our discussion it is required by
spacetime supersymmetry.  

The topological string theory representation
of the superpotential allows us to write the full
worldsheet instanton contribution to the
CFT correlator \threept.
First, note that while
we have discussed $A_i$ as a harmonic form,  
we can modify it by adding a BRST-trivial piece to give it
support only in an arbitrarily  small neighborhood around
a two-cycle $\beta_i$ which is Poincar\'e
dual to $\gamma_i$.
The result is as follows.
Denote by $d^{\{n_a\}}_{\{m_l\}}(i,j,k)$ the number of 
holomorphic maps from a disc
to $M$ where the image
$D \subset M$ has the following properties:\foot{As  
with ``numbers'' of rational curves in mirror symmetry,
the correct notion of $d$ when there are families of discs and/or
including multiple covers would require much further discussion; we
will be content here to be schematic.  A proposal for the multiple
cover contribution has recently been worked out by H. Ooguri and
C. Vafa \ov.}
\smallskip
\noindent
i)~~$[\partial D] ~=~ \sum_{l} m_{l}~\gamma_l$.
\smallskip
\noindent
ii)~~The vertex operators $V^{i,j,k}$ are mapped
in cyclic order to intersections of $\gamma=\p D$
with $\beta_{i,j,k}$ respectively.\foot{One has to be
careful if two vertex operators correspond to the
same cycle.  The support of 
$A$ can be made arbitrarily small but finite.  In this
way nonzero contributions still generically come from
the vertex operator insertions mapping to different
points in $\gamma$.}
\smallskip
\noindent
iii)~~ $D - \sum_{l} m_{l} D_{l}$, 
which is a closed two cycle in $M$, is
in the homology class
$\sum_{a} n_{a} K_{a}$.

\noindent
Then, the three-point function receives a contribution
\eqn\wthree{\eqalign{
	\langle
	V^{i}_F V^j_\phi V^k_\phi \rangle  
	~\sim~ \sum_{m_l, n_a \geq 0} 
	~&\left(\int_{\partial D} \theta^{i}\right)
	~\left(\int_{\partial D}
	\theta^{j}\right)~\left(\int_{\partial D} \theta^{k}\right)
	\times\cr
	&\times d^{\{n_a\}}_{\{m_l\}}(i,j,k) 
	\prod_{l=1}^{b^{1}(\Sigma)}
	e^{-m_l (w_l + i a_l)/\ap} 
	\prod_{a=1}^{h^{1,1}(M)} e^{-n_a t_a}}}
from disc instantons, where $t_a$ denotes the
integral of the K\"ahler form  
over $K_a$ (and for simplicity we are setting the closed string
background $B$-field to zero).  Although we have
mostly used the language of the topological theory in deriving this
result, it also holds for  
the three-point function in the physical theory. 

The same kind of instanton sum
also appears in \bundlemir, where 
the interpretation in terms of
a superpotential for wrapped branes (and in particular 
the fact that these effects serve
to obstruct the deformations of branes wrapped on special 
Lagrangian cycles) was not discussed.

\vfill\eject
\bigskip{\noindent{\it Coupling to closed string background fields}}

It is clear from the form of the three-point 
functions \wthree\ that the superpotential
depends on the closed string background K\"ahler 
moduli, which enter through
the worldsheet instanton action $e^{-t_a}$.  
The dependence of the disc
instanton generated superpotential on K\"ahler moduli, 
and the fact that it does
$\it not$ depend on the background complex 
structure moduli in the IIA theory,
is consistent with the nonrenormalization result of \bdlr.

We can directly probe the dependence of the superpotential on closed
string moduli by computing the CFT correlator
\eqn\closop{\langle V_{K}^{(-1,-1),a} V_{F}^{(0),j}\rangle}
where $V_{K}^{(-1,-1),a}$ is the vertex operator for 
a closed string K\"ahler
deformation.  Again, the vertex operators in \closop\ 
can absorb the fermion zero modes which are
present in an instanton background.
In fact, the ``mirror'' couplings of open strings to
background complex moduli in the superpotential 
generically exist at tree level in the
B-model \bdlr\ -- this is clear from the examples 
of \S2.2, where a small
perturbation of complex structure can obstruct
families of holomorphic curves.
The couplings \closop\ must then similarly exist, 
but due to Peccei-Quinn symmetries
they should arise at the non-perturbative level in 
both the closed and open string
worldsheet instanton expansions.

$V_{K}^{a}$ represents an 
integral (1,1) form $\omega_a$ which could
be used to perturb the K\"ahler form of $M$.  
We can choose $\omega_a$ to have
support only infinitesimally near the four-cycle $L^{a} 
\subset M$ Poincare dual to $K_a$.  
Then, \closop\ has the expansion
\eqn\anotsum{
\eqalign{
	\langle V_{K}^{(-1,-1),a} 
	V_{F}^{(0),j}\rangle \sim
	&\sum_{m_l, n_b \geq 0} \left( \int_{D} \omega_a \right) 
	\left( \int_{\partial D} \theta^{j} \right) \times\cr
	&\times
	d^{\{n_b\}}_{\{m_l\}}(a,j) \prod_{l=1}^{b^{1}(\Sigma)} 
	e^{-m_l(w_l + ia_l)/\ap}
	\prod_{b=1}^{h^{1,1}(M)} e^{-n_b t_b}
}}
where $d^{\{n_b\}}_{\{m_l\}}(a,j)$ counts the number of 
holomorphic maps to discs
$D \subset M$ which pass through $L^a$ at the 
insertion point of $V^{a}_K$ and
$\beta_j$ at the insertion point of $V^{j}_{F}$, 
and which in addition have
$[\partial D] = \sum_l m_l \gamma_l$ and 
$[D - \sum_l m_l D_l] = \sum_b n_b K_b$.

\newsec{Discussion}

Space-filling D-branes wrapping supersymmetric 
cycles in Calabi-Yau manifolds provide one
of the most natural classes of ${\cal N}=1$ 
supersymmetric models in string theory, and
are attractive as concrete realizations of ``brane world'' 
scenarios.  In this paper, we
have shown that the theories which arise from D6 branes 
wrapping supersymmetric three-cycles
are in many ways analogous to heterotic (0,2) models.  
In particular, although they are
supersymmetric to all orders in $\ap$, nonperturbative worldsheet effects can
generate superpotentials and, perhaps, break supersymmetry.

These models differ from heterotic theories, however, in that
mirror symmetry provides a dual description where the non-perturbative 
superpotential is computable at tree level in
sigma model perturbation theory.  This should be a powerful tool: 
most known dualities of ${\cal N}=1$ models, like 
heterotic/F-theory duality, relate instanton
computations to other instanton computations (with 
worldsheet instantons mapping to
euclidean wrapped branes of various sorts \edinst).  
The present situation is considerably
rosier, and it will be very interesting to exploit this to sum up instantons
in this class of ${\cal N}=1$ string vacua.

The cases discussed in \S2.2\ (on the B-model side) 
should provide ideal test cases.  In each case, one 
can realize (on a 5-brane wrapping a holomorphic curve) a theory
with massless chiral fields, constrained by a higher order 
superpotential.  The mirror D6 theory should provide 
us with an example of a brane wrapping a
supersymmetric three-cycle $\Sigma$ with
$b_{1}(\Sigma) > 0$, but without a moduli space of the expected dimension.
By the nonrenormalization theorem of
\S2.1, the moduli space on the A-model side
must be lifted by a disc instanton generated superpotential.
Work to explicitly construct the mirror cycles, and 
compute the relevant superpotentials, is under way \toapp.
Note that nonperturbative superpotentials which obstruct
deformations of branes wrapped on three-cycles can resolve the puzzle 
for mirror symmetry raised by
Thomas in \thomas.  On the other hand, we expect e.g. the
supersymmetric $T^3$ used in \syz\ to derive mirror symmetry will
survive instanton corrections.  In the mirror picture this is
obvious (since deformations of a point are unobstructed), and
in the direct analysis presumably any holomorphic discs 
with boundary on the $T^3$ would come in families and cancel 
in their contribution to the superpotential.

In the regime where there are ``small'' holomorphic discs, 
new interesting phenomena should
also occur.  For instance, there are arguments in the 
mathematics literature that in some cases the classical
moduli spaces of special Lagrangian three-cycles will be 
manifolds with boundary (see \S5 of \joyceslag).  
This cannot be the case for physical applications 
of the sort we have discussed, involving
wrapped branes in string theory.
The moduli space (including Wilson line degrees of freedom) is that of 
a 4d ${\cal N}=1$ supersymmetric
D-brane field theory.  Assuming supersymmetry isn't broken, the quantum 
moduli space of supersymmetric
ground states must be a K\"ahler manifold; there is no known dynamics 
that can create boundaries at codimension one in the moduli space 
of 4d ${\cal N}=1$ supersymmetric theories.  The argument of \joyceslag\
involves the fact that a holomorphic disc with boundary in the 
three-cycle is becoming very small;
therefore, it is likely that some analogue of the phenomena 
discussed in \edphases\ is occurring.  Just as one
can use theta angles to go around the boundaries of the classical 
K\"ahler cone and find intrinsically
stringy Landau-Ginzburg phases of Calabi-Yau 
compactifications, it seems likely that one
can use Wilson lines to go around the would-be boundary of moduli space
discussed in \joyceslag\ and find new, ``quantum'' supersymmetric
three-cycles.

\bigskip 
\centerline{\bf Acknowledgements}
\smallskip
We would like to thank P. Aspinwall, D. Brace, H. Clemens,
D.-E. Diaconescu, M. Douglas, S. Gukov, A. Klemm,
G. Moore, D. Morrison, T. Pantev, R. Plesser, E. Silverstein and C. Vafa
for interesting discussions about related subjects.
This work was initiated at the Aspen Center for
Physics.  S. Kachru and A. Lawrence would like to acknowledge the
kind hospitality of N. Seiberg and
the Institute for Advanced Study while the bulk of 
this work was carried out.
S. Kachru was supported in part by the generosity of the
Ambrose Monell Foundation, an A.P.
Sloan Foundation Fellowship and a DOE OJI Award;
S. Katz was supported in part by NSA grants MDA904-98-1-0009 and
MDA904-00-1-0052;
A. Lawrence was supported in part by the
DOE under contract DE-AC03-76F00515, and in part by
the NSF via grant PHY94-07194 through the Institute for Theoretical
Physics; and J. McGreevy was supported in
part by the Department of Defense NDSEG Fellowship program.

\bigskip

\listrefs

\end